\documentstyle[prc,aps,epsf,epsfig,amssymb,amsbsy,amstext,amsfonts]{revtex}
\begin{document}
\title{ The lightest scalar nonet\footnote{
Invited talk to be published in the proceedings of the workshop
       {\it Chiral fluctuations in hadronic matter},
 Orsay, September 16-28, 2001             }
       }
\author{Nils A. T\"ornqvist}
\address{Physics Department \\
University of Helsinki \\
POB 63, FIN--00014,  Finland}
\maketitle
\begin{abstract}
First I review some previous work
on the lightest scalars below 1.5 GeV, and how these scalars
can be understood as unitarized $q\bar q$,
$q\bar q q\bar q$ or meson-meson nonet states. The bare
scalars are  strongly distorted  by hadronic mass shifts, and the
lightest $I=0$ state becomes a very broad resonance of mass and width of about 500 MeV.
 This is the $\sigma$ meson required by models based on linear realization
of chiral symmetry.
Recently the light $\sigma$ has
clearly been observed in $D\to\sigma\pi\to3\pi$ by the E791
experiment at Fermilab and I  discuss how this decay channel can be
predicted in a Constituent Quark Meson Model (CQM), which
incorporates heavy quark and chiral symmetries.

At the end I discuss the
likely possibility that there are in fact two light scalar nonets,
such as one mainly meson-meson (or 4-quark) nonet and one $q\bar q$ nonet.
I point out that an interesting approximate description  of this could
be modelled by starting with two coupled linear sigma models.
After gauging the overall symmetry one of these could
be looked upon as the "Higgs sector of strong interactions", and the lightest
scalar nonet becomes the corresponding  Higgs nonet.
\end{abstract}

\def \gam {\frac{ N_f N_cg^2_{\pi q\bar q}}{8\pi} }
\def \gam {\frac{ N_f N_cg^2_{\pi q\bar q}}{8\pi} }
\def \gamm {N_f N_cg^2_{\pi q\bar q}/(8\pi) }
\def \be {\begin{equation}}
\def \ba {\begin{eqnarray}}
\def \ee {\end{equation}}
\def \ea {\end{eqnarray}}
\def \gap {{\rm gap}}
\def \gapp {{\rm \overline{gap}}}
\def \gappp {{\rm \overline{\overline{gap}}}}
\def \im {{\rm Im}}
\def \re {{\rm Re}}
\def \Tr {{\rm Tr}}
\def \P {$0^{-+}$}
\def \S {$0^{++}$}
\def \uu {$u\bar u$}
\def \dd {$d\bar d$}
\def \ss {$s\bar s$}
\def \qq {$q\bar q$}
\def \qqq {$qqq$}
\def \lsm {L$\sigma$M}
\def \sig {$\sigma$}
\def \gam {\frac{ N_f N_cg^2_{\pi q\bar q}}{8\pi} }
\def \gamm {N_f N_cg^2_{\pi q\bar q}/(8\pi) }
\def \be {\begin{equation}}
\def \ba {\begin{eqnarray}}
\def \ee {\end{equation}}
\def \ea {\end{eqnarray}}
\def\bea{\begin{eqnarray}}
\def\eea{\end{eqnarray}}
\def \gap {{\rm gap}}
\def \gapp {{\rm \overline{gap}}}
\def \gappp {{\rm \overline{\overline{gap}}}}
\def \im {{\rm Im}}
\def \re {{\rm Re}}
\def \Tr {{\rm Tr}}
\def \P {$0^{-+}$}
\def \S {$0^{++}$}
\def\zpp{$0^{++}$}
\def\fz{$f_0(980)$}
\def\az{$a_0(980)$}
\def\Kz{$K_0^*(1430)$}
\def\fzz{$f_0(1370)$}
\def\fzzz{$f_0(1200-1300)$}
\def\azz{$a_0(1450)$}
\def\ss{$ s\bar s $}
\def\uu{$u\bar u+d\bar d$}
\def\qq{$q\bar q$}
\def\KK{$K\bar K$}
\def\sig{$\sigma$}
\def\lsim{\;\raise0.3ex\hbox{$<$\kern-0.75em\raise-1.1ex\hbox{$\sim$}}\;}
\def\gsim{\raise0.3ex\hbox{$>$\kern-0.75em\raise-1.1ex\hbox{$\sim$}}}

\section{Introduction}
This talk is partly based on published papers~\cite{NAT1,NAT2,gatto,deandrea,lsm} on
 the lightest scalars including a few new comments, and partly on
some work under progress.
First I shall
discuss the evidence for the light \sig\ and discuss shortly the linear sigma model,
 and explain how one can
understand the controversial light scalar mesons with a unitarized
quark  model, which includes most well established
theoretical constraints:
{\it (i)} zeroes as required by chiral symmetry,
{\it (ii)} all light two-pseudoscalar (PP) thresholds with flavor symmetric couplings in a coupled channel framework
{\it (iii)} physically acceptable analyticity, and
{\it (iv)}  unitarity.

  A unique
feature of such a model is that it simultaneously describes a  whole scalar
nonet and one obtains a good representation of a large
set of relevant data, with a few parameters,
which all have a  clear physical interpretation.
Also consistency with unitarity can require that when the effective coupling becomes large enough
one can have twice as many poles in the output spectrum,
 than the $q\bar q $-like poles one puts in as Born terms.
The new poles can then be interpreted as being mainly meson-meson bound states, but strongly
mixed with $q\bar q$ states.

Then I
discuss the recently measured $D\to\sigma\pi\to 3\pi$
and $D_s\to f_0(980)\pi\to 3\pi$ decays, where
the \sig , respectively $f_0(980)$, is clearly seen as the dominant peak, and point out that these
decays rates can be understood in a rather general model for the weak matrix elements. This
indicates that the broad \sig (600) and the $f_0(980)$ belong to the same multiplet.

Finally the experimental fact that there seems to be too many light scalar
mesons, presumably two nonets, one in the 1 GeV region  (a meson-meson nonet)
and another one near 1.5 GeV (a  $q\bar q$  nonet),
 requires a new effective model for the light
scalar spectrum.

I argue  that two coupled linear sigma models may provide
a first step for an understanding of such a proliferation of 18 light scalar states.
After gauging the overall symmetry one  could then look at
the lightest scalars as Higgs-like bosons for the nonperturbative low energy
strong interactions.

\section{The problematic scalars and the existence of the \sig }

The interpretation of the nature of lightest scalar mesons has
been controversial for long. There is no general agreement on
where are the  $q\bar q$ states, is there a glueball among the
light scalars, are some of the too numerous scalars multiquark, $K\bar K$
or other meson-meson bound states?
These are fundamental questions of great importance in particle
physics. The mesons with vacuum quantum numbers are known to be
crucial for a full understanding of the symmetry breaking
mechanisms in QCD, and presumably also for confinement.

As for the light and broad $\sigma$ near 600 MeV, all authors do not even agree on
its existence as a fundamental hadron, although the number of
supporters is growing rapidly.

\begin{figure}
\epsfxsize=17 cm
\centerline{\epsffile{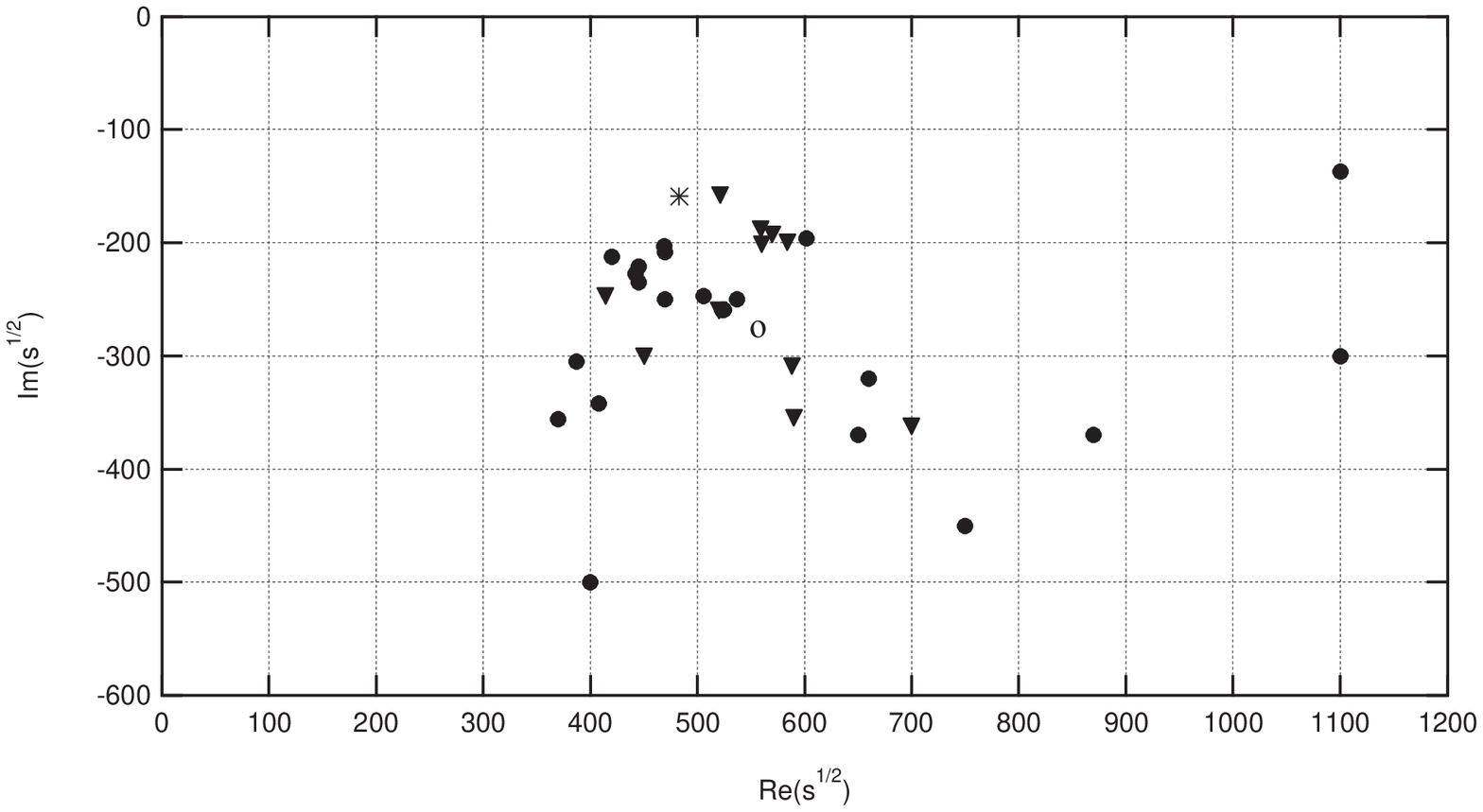}} \noindent {\bf Fig. 1}
 - {
The pole positions of the $\sigma$
resonance, as listed by the PDG (year 2000)
under $f_0(400-1200)$ or $\sigma$ (filled circles), plotted in the complex
energy plane (in units of MeV). The triangles represent the mass
and width parameters (plotted as $m-i\Gamma /2$), which were
reported at a Kyoto  meeting in June 2000. I could not here distinguish between
pole and Breit-Wigner parameters. The star is the $m-i\Gamma /2$
point obtained from the recent E791 experiment on
$D\to\sigma \pi\to 3\pi$ ($m_\sigma=478$ MeV, $\Gamma_\sigma =
324$ MeV) while the open circle is that obtained by CLEO
for $\tau\to\sigma\pi\nu \to 3\pi\nu$.
}
\end{figure}

A light scalar-isoscalar meson (the \sig ), with a mass of twice
the constituent  $u,d$ quark mass, or $\approx 600$ MeV, coupling
strongly to $\pi\pi$ is of importance in all
Nambu--Jona-Lasinio-like (NJL-like) models for dynamical breaking
of chiral symmetry. In these models the $\sigma$ field obtains a
vacuum expectation value, i.e., one has a \sig-like
condensate in the vacuum, which is crucial for the understanding
of all hadron masses, as it explains in a simple way the
difference between the light constituent and chiral quark mass.
 Then most of the nucleon mass is generated by its coupling to
the $\sigma$, which acts like an effective Higgs-like
 boson for the hadron spectrum.

In Fig. 1 I have plotted with filled circles the results of 22
different analyses on the \sig\ pole position, which are included
in the  2000 edition of the Review of Particle Physics
\cite{pdg2000} under the entry $f_0(400-1200)$ or \sig.  Most of
these find a \sig\ pole position near 500-i250 MeV.

Also, at  a recent meeting in Kyoto \cite{kyoto} devoted to the
$\sigma$, many groups reported preliminary analyzes, which find
the \sig\ resonance parameters in the same region. These are
plotted as triangles in Fig. 1. Here it was not possible to
distinguish between Breit-Wigner parameters and pole positions,
which of course can differ by several 100 MeV for the same data.
 It must also be noted that many of the triangles in Fig. 1 rely on the same
raw data and come from preliminary analyzes not yet published.

I also included in Fig. 1 (with a star) the \sig\ parameters
obtained from the recent E791 Experiment at Fermilab \cite{E791},
where 46\% of the $D^+\to3\pi$ Dalitz plot is $\sigma\pi$.  The
open circle in the same figure represents the \sig\ parameters
extracted from the CLEO analysis of $\tau\to\sigma\pi\nu\to
3\pi\nu$ \cite{CLEO}. There is also a very clear, although still
preliminary, signal for a light \sig\ (Breit-Wigner mass=$390^{+60}_{-36}$ and width=
$282^{+77}_{-50}$) in a BES experiment\cite{wu}
on J/$\psi\to\sigma\omega\to\pi\pi\omega$.

\section{The NJL and the linear sigma model}

The NJL model is an effective theory which is believed to be
related to QCD at low energies, when one has integrated out the
gluon fields. It involves a linear realization of chiral symmetry.
After bosonization of the NJL model one finds essentially the
linear sigma model (\lsm ) as an approximate effective theory for
the scalar and pseudoscalar meson sector.

About  30 years ago Schechter and Ueda \cite{u3u3}  wrote down the
$U3\times U3$ \lsm\ for the meson sector involving a scalar and a
pseudoscalar nonet. This (renormalizable) theory has only 6
parameters, out of which 5 can be fixed by the pseudoscalar masses
and decay constants ($m_\pi,\ m_K, \ m_{\eta^\prime},\ f_\pi, \
f_K$). The sixth parameter for the OZI rule violating 4-point
coupling must be small. One can then predict, with no free
parameters, the tree level scalar masses \cite{lsm}, which turn
out to be not far from the lightest experimental masses, although
the two quantities (say Lagrangian mass vs. second sheet pole mass))
are \underline{not}  the same thing, but
can differ for the same model and data by well over 100 MeV.

The important thing is that the scalar masses are predicted to be
near the lightest experimentally seen scalar masses, and not in
the 1500 MeV region where many authors want to put the lightest
$q\bar q$ scalars. The \sig\ is predicted \cite{lsm} at 620 MeV
with  a very large width ($\approx 600$ MeV), which well agrees
with Fig. 1. The $a_0(980)$ is predicted at 1128 MeV,  the
$f_0(980)$ at 1190 MeV, and the $\kappa$ or $K^*_0(1430)$ at 1120 MeV, which
is still surprisingly good considering that loops or unitarity   effects must be large
as we discuss in the next section.

\section{ Understanding the S-waves within a unitarized quark model (UQM)}

A few years ago  I presented fits to the $K\pi$, $\pi\pi$ S
-waves and to the  \az\ resonance peak in $\pi\eta$\cite{NAT1,NAT2}.
In order to understand one of the main points of that work it is sufficient
to look at the partial
wave amplitude (PWA) in the case of one input resonance, such one
 I=1/2 ($K^*_0$) resonance, or  one I=1 ($a_0(980)$) resonance. It
 can  for $\pi\eta\to\pi\eta$
 be written simply as:
\begin{equation} A(s)=-
\frac{Im\Pi_{\pi\eta}(s)}{[m_0^2+Re\Pi (s)-s +iIm\Pi (s)]},
\end{equation}
where: \bea \label{PWA} Im\Pi (s)&=&\sum_i Im\Pi_i(s) \label{impi}
=-\sum_i \gamma_i^2(s-s_{A,i})\frac{k_i}{\sqrt s}e^{-k_i^2/k_0^2}
           \theta(s-s_{th,i})\ ,\nonumber \\
Re\Pi (s)&=&\frac 1\pi{\rm P.V.}\int^\infty_{s_{th,1}} \frac{Im\Pi
(s)}{s'-s} ds' \ . \nonumber \eea Here the coupling constants
$\gamma_i$ are related by flavour symmetry and OZI rule, such that
there is only one over all parameter $\gamma$. The $s_{A,i}$ are
the positions of the Adler zeroes, which  are  near $s=0$. Eq. (1)
can be looked upon as a more general Breit-Wigner form, where the
mass parameter is replaced by an $s$-dependent function, ``the
running mass" $m_0^2+ Re\Pi (s)$. This $s$-dependent mass coming from Re$\Pi (s)$ is  crucial
for understanding the scalars. Because of the S-wave thresholds, giving cusps,
 and the large effective coupling this cannot be neglected.

In the flavourless channels the situation is  more
complicated than in Eq. (1) since one has two $I=0$
states, requiring a two dimensional  mass matrix with $s$-dependent mass matrix and mass mixing.
Note that the sum runs over all light PP
thresholds, which means three for the \az : $\pi\eta ,\ K\bar
K,\pi\eta'$ and three for the \Kz : $ K\pi ,\ K\eta ,\ K\eta' $,
while for the $f_0$'s there are five channels: $\pi\pi ,K\bar K ,\
\eta\eta ,\ \eta\eta' ,\ \eta'\eta'$.

In Figs. 2 and 3 I show the running mass,
$m_0^2+Re\Pi(s)$, and the width-like function, $-Im\Pi(s)$, for
the I=1/2 and I=1 channels. The crossing point of the running mass with $s$
gives the $90^\circ$  mass of the \az.
 The magnitude of the \KK\ component in the \az\ is determined by
$-\frac d{ds}Re\Pi(s)$, which is large in the resonance region
just below the \KK\ threshold. These functions fix the PWA of Eq. (1).

In Ref.~\cite{NAT2} the \sig\ was missed because only poles
nearest to  the physical region were looked for, and the
possibility of the resonance doubling phenomenon, discussed below,
was overlooked. Only a little later we realized with Roos
\cite{NAT1} that two resonances (\fz\ and $ f_0(1370)$) can emerge
although only one $s\bar s$ bare state is put in. Then we had to
look deeper into the second sheet and found the broad \sig\ as the
dominant singularity at low mass.

In fact, it was pointed out by Morgan and Pennington~\cite{morgan}
that for each \qq\ state there are, in general, apart from the
nearest pole, also  image poles, usually located far from the
physical region. As explained in more detail in Ref.~\cite{NAT1},
some of these can (for a large enough coupling and sufficiently
heavy threshold) come so close to the physical region that they
make new resonances. And, in fact, there are more than four
physical poles  with different isospin, in the output spectrum of
the UQM model, although only four bare states, of {\it the same
nnonet},  are put in!. The \fz\ and the \fzz\ of the
model thus turn out to be  two manifestations of the same \ss\
state. Similarly the \az\ and the  \azz\ could be two manifestations of the $u\bar d$ state.

\begin{figure}
\epsfxsize=15cm
\centerline{\epsffile{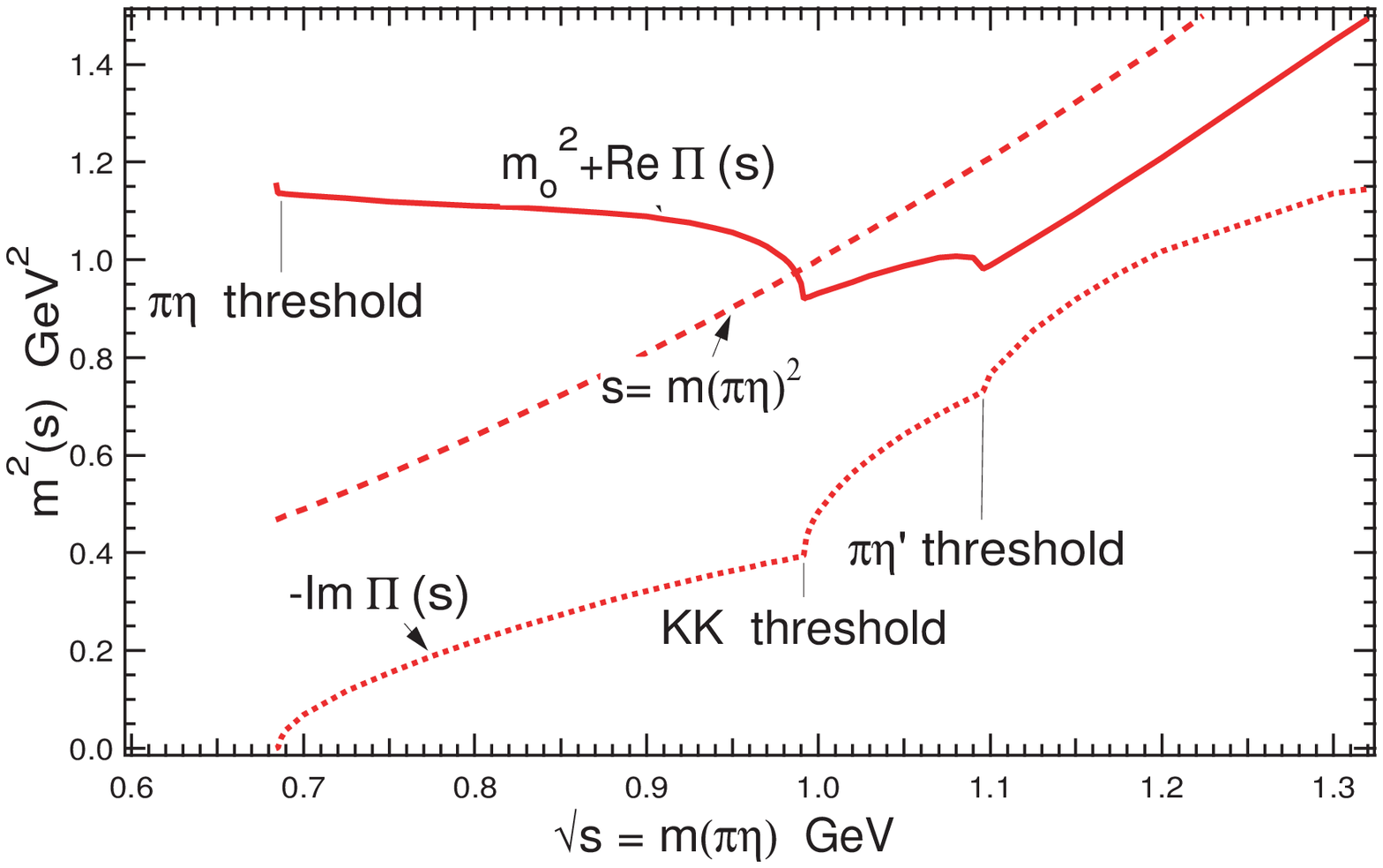}} \noindent {\bf Fig. 2}
 - {The running mass $m_0+ Re\Pi(s)$ and $Im \Pi (s)$ of
the $a_0(980)$. The strongly dropping running mass at the
$a_0(980)$ position, below the $K\bar K$ threshold contributes to
the narrow shape of the peak.}
\end{figure}

\begin{figure}
\epsfxsize=14cm
\centerline{\epsffile{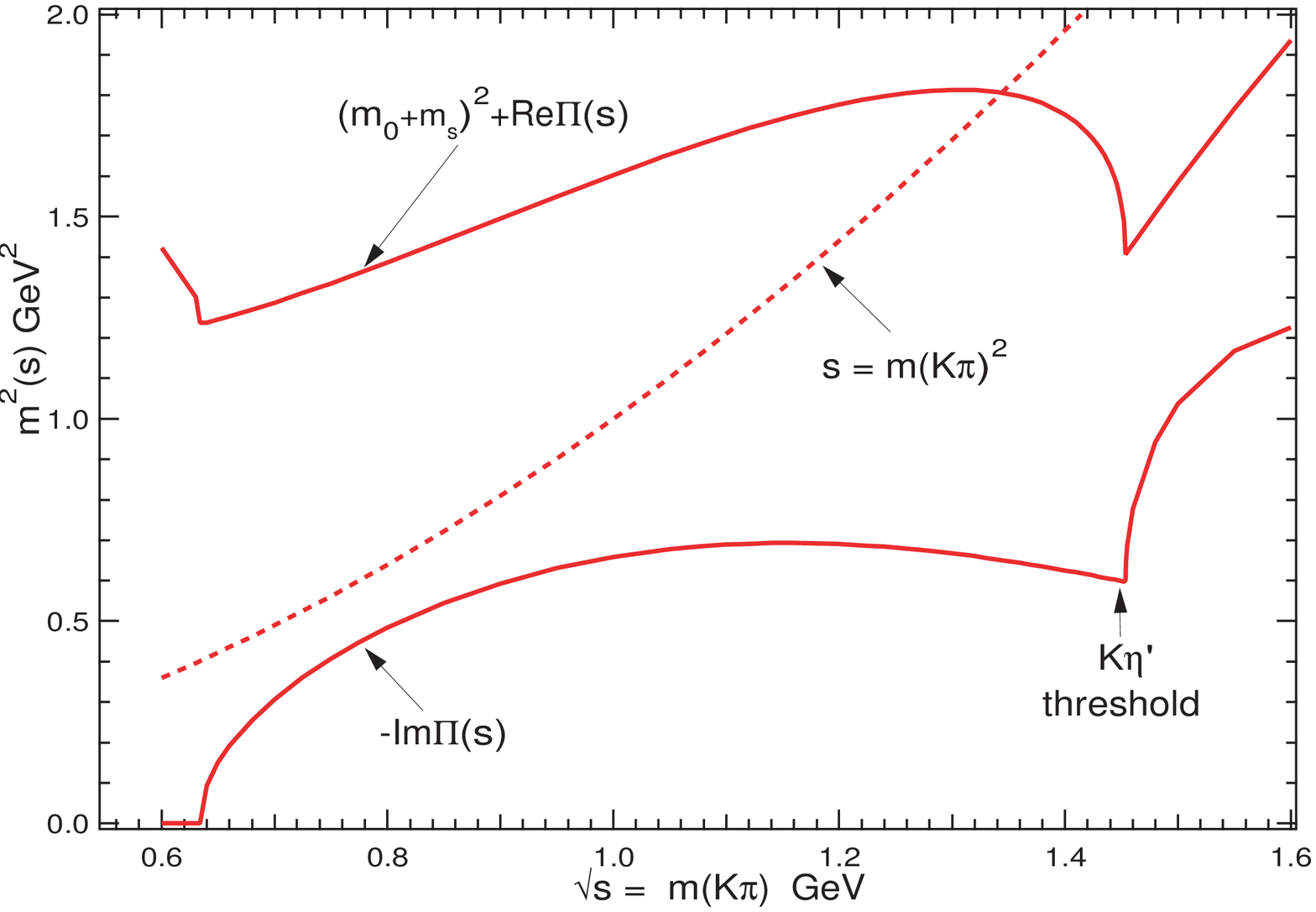}} \noindent {\bf Fig. 3}
 - {The running mass and width-like function $-Im\Pi(s)$
for the $K^*_0(1430)$. The crossing of $s$ with the running mass
gives the 90$^\circ$ phase shift mass, which roughly corresponds
to a naive Breit-Wigner mass, where the running mass is put
constant.
}
\end{figure}
\newpage

This phenomenon  can  manifest itself by two
crossings  with the running mass $m_0^2+ Re\Pi(s)$,
one near the threshold and another at higher mass. Then each crossing is
related to a different pole. This happens if the
coupling is strong enough and one would
generate an extra bound state pole, below the threshold.
(For $s\bar\to K\bar K$ one is close to that situation,
 and in the case of $u\bar s \to K\pi$ a 50\% larger coupling
 physical coupling could bind $K\pi$). That bound state
pole would then be a true "2-meson bound state" of the two pseudoscalars
making up the threshold in question. The "binding mechanism" is then
generated by the sum of $s$-channel loops required by unitarity.
Such binding is of course, in general, related to crossed channel exchanges,
which in the UQM are simulated by the form factor.

Although the details of any modelling can be critisised, the conclusion remains:
{\it Strong enough couplings can generate new bound states or resonances not present in the input
or in the Born terms represented by an effective Lagrangian}.

Only after  realizing that this resonance doubling is
important the light and broad \sig\ was found in the model.
(See Ref. \cite{NAT1} for details).
Another important effect that the model can explain is the large
mass difference between the $a_0$ and $K^*_0$. Because of this
large mass splitting many authors argue that the $a_0(980)$ and
$f_0(980)$ are not \qq\ states, since in addition to being very
close to the \KK\ threshold, they are much lighter than the first
strange scalar, the  $K^*_0(1430)$. Naively one expects a mass
difference between the strange and nonstrange meson to be of the
order of the strange-nonstrange quark mass difference, or a little
over 100 MeV.

Figs. 2 and 3 explain why  one can easily understand this large
$K^*_0(1430)-a_0(980)$ mass splitting as a secondary effect of the
large pseudoscalar mass splittings, and because of the large mass
shifts coming from the loop diagrams involving the PP thresholds.
If one puts Figs. 2 and 53 on top of each other one  sees that  the
3 thresholds $\pi\eta,\ K\bar K,\ \pi\eta$ all lie relatively
close to the $a_0(980)$, and all 3 contribute to a large mass
shift. On the other hand, for the $K^*_0(1430)$, the $SU3_f$
related thresholds ($K\pi,\ K\eta'$) lie far apart from the
$K^*_0$, while the $K\eta$ nearly decouples because of the
physical value of the pseudoscalar mixing angle.

This large mass of the $K^*_0(1430)$ is also one of the reasons
why several authors want to have a lighter strange meson, the
$\kappa$, near 800 MeV. Cherry and Pennington \cite{cherry}
 have strongly argued against its existence. But the E791
experiment  see some evidence for such a
light $\kappa$ in  $D^+\to K^-\pi^+\pi^+$. Here
the signal is much less evident than the $\sigma$ in $D\to 3\pi$,
but the $\kappa$ improves the $\chi^2$ in the
region dominated by the $K^*(890)$. One should  try a more
sophisticated Breit-Wigner amplitude for the S-wave, as that in
Eq. (1), before one makes more definite statements about the
$\kappa$ It  could be something like a virtual bound $K\pi$
state.
\section{$D\to\sigma\pi\to 3\pi$ and $D_s\to f_0(980)\pi\to 3\pi$}

The recent experiments studying charm decay to light hadrons are
opening up a new experimental window for understanding light meson
spectroscopy and especially the controversial scalar mesons, which
are copiously produced in these decays.

In particular we refer to the E791 study of the $D\to 3\pi$ decay
\cite{E791} where it is shown how adding an intermediate scalar
resonance with floating mass and width in the Monte Carlo program
simulating the Dalitz plot densities, allows for an excellent fit
to data provided the mass and the width of this scalar resonance
are $m_\sigma\simeq 478$ MeV and $\Gamma_\sigma\simeq 324$ MeV.
This resonance is a very good candidate for the $\sigma$. To check
this hypothesis we adopt the E791 experimental values for its mass
and width and using a Constituent Quark Meson Model (CQM) for
heavy-light meson decays \cite{rass} we compute the
$D\to\sigma\pi$ non-leptonic process via {\it factorization}
\cite{WSB}, taking the coupling of the $\sigma$ to the light
quarks from the linear sigma model \cite{volkoff}. In such a way
one is directly assuming that the scalar state needed in the E791
analysis could be the quantum of the $\sigma$ field of the linear
sigma model. According to the CQM model and to factorization, the
amplitude describing the $D\to\sigma\pi$ decay can be written as a
product of the semileptonic amplitude $\langle
\sigma|A^\mu_{(\bar{d}c)}(q)|D^+\rangle$, where $A^\mu$ is the
axial quark current, and $\langle \pi|A_{\mu(\bar{u}d)}(q)|{\rm
VAC}\rangle$. The former is parameterized by two form factors,
$F_1(q^2)$ and $F_0(q^2)$, connected by the condition
$F_1(0)=F_0(0)$, while the latter is governed by the pion decay
constant $f_\pi$. As far as the product of the two above mentioned
amplitudes is concerned, only the form factor $F_0(q^2)$ comes
into the expression of the $D\to\sigma\pi$ amplitude. Moreover we
need to estimate it at $q^2\simeq m_\pi^2$, that is the physically
realized kinematical situation. The CQM offers the possibility to
compute this form factor through two quark-meson 1-loop diagrams
that we call the {\it direct} and the {\it polar} contributions to
$F_0(q^2)$. These quark-meson loops are possible since in the CQM
one has effective vertices (heavy quark)-(heavy meson)-(light
quark) that allow us to {\it compute} spectator-like diagrams in
which the external lines represent incoming or outgoing heavy
mesons while the internal lines are the constituent light quark
and heavy quark propagators.

In Figs. 2 and 3 of Ref.\cite{gatto} we show respectively the {\it direct} and the
{\it polar} diagrams for the semileptonic amplitude $D\to\sigma$,
the former being characterized by the axial current directly
attached to the constituent quark loop, the latter involving an
intermediate $D(1^+)$ or $D(0^-)$ state. These two diagrams are
computed with an analogous technique and one finally obtains a
determination of the direct and polar form factors $F_0^{\rm dir,
pol}(q^2)$. The extrapolation to $q^2\simeq m_\pi^2\simeq 0$ is
safe for  the direct form factor while is not perfectly under
control for the polar form factor since the latter is more
reliable at the pole $q^2\simeq m_P^2$, $m_P$ being the mass of
the intermediate state. We take into account the
uncertainty introduced by this extrapolation procedure and
signaled by the fact that we find $F_0^{\rm pol}(0)\neq F_1^{\rm
pol}(0)$ (computing $F_0$ from the polar diagram with $0^-$
intermediate polar state and $F_1$ from that with intermediate
$1^+$ state). Our estimate for $F_0(0)=F_0^{\rm pol}(0)+F_0^{\rm
dir}(0)=0.59\pm 0.09$ is in reasonable agreement with an estimate
of $F_0(m_\pi^2)=0.79\pm 0.15$ carried out in \cite{dib} using the
E791 data analysis and a Breit-Wigner like approximation for the

The meson-quark loops are computed substituting
the meson vertices with the heavy meson field expressions found by
Heavy Quark Effective Theory (HQET) (since CQM incorporates heavy
quark and chiral symmetries) and the quark lines with the
propagators of the heavy and light constituent quarks. The light
constituent mass $m$ is fixed by a NJL-type gap equation that
depends on $m$, and on two cutoffs $\Lambda$ and $\mu$ in a proper
time regularization scheme for the diverging integrals. The
ultraviolet cutoff $\Lambda$ is fixed by the scale of chiral
symmetry breaking, $\Lambda_\chi\simeq 4\pi f_\pi$, and we
consider $\Lambda=1.25$ GeV. The remaining dependence of $m$ on
the choice of the infrared cutoff $\mu$ has an expression similar
to that of a ferromagnetic order parameter, $m(\mu)$ being
different from zero for $\mu$ values smaller than a particular
$\mu_c$, and zero for higher values. When $\mu$ ranges from $0$ to
$300$ MeV, the value of $m$ is almost constant, $m=300$ MeV,
dropping for higher $\mu$ values. A reasonable light constituent
quark mass is certainly $300$ MeV and this clearly leaves a $300$
MeV open window for choosing the infrared cutoff. Enforcing the
kinematical condition for the meson to decay to its free
constituent quarks, which must be possible since the CQM model
does not incorporate confinement, requires $\mu\simeq m$
\cite{rass}. Therefore we pick up the $\mu=300$ MeV value. The
results are quite stable against $10-15\%$ variations of the UV
and IR cutoffs.

The CQM semileptonic $D\to\sigma$ transition amplitude is
represented by the loop integrals associated to the direct and to
the polar contributions. The result of the integral computations
must then be compared with the expression for the hadronic
transition element  $\langle \sigma |A|D\rangle$ and this allows
to extract the desired $F_{0,1}$ form factors. An estimate of the
weight of $1/m_c$ corrections can also be taken into account
\cite{gatto}.

This computation indicates that the scalar resonance described in
the E791 paper can be consistently understood as the $\sigma$ of
the linear sigma model. Of course a calculation such as the  one
here described calls for alternative calculations and/or
explanations of the E791 data for a valuable and useful comparison
of point of views on the $\sigma$ nature.

A similar computation\cite{deandrea} was performed for the related
process $D_S\to f_0(980)\pi\to 3\pi$ with the same model. The
agreement with the data  indicates that the two resonances
the $\sigma$ and the $f_0(980)$ belong to a similar flavour multiplet.

\section{Two coupled  linear sigma models for the two scalar nonets}

As we have seen in the previous discussion there seems to be a
proliferation of light scalar mesons.
Let us assume\cite{fut} that we have two
two light nonets of scalar mesons  below about, say 1.7 GeV, one of which
is the  $q\bar q$ nonet expected from QCD or the quark model,
 while the other is e.g. meson-meson
bound states, but also in a nonet.
In order to have a realistic effective model at low energies for
the scalars we need an effective  chiral quark model, which includes all
scalars and pseudoscalars, and where the chiral symmetry is broken
by the vacuum expectation values of the scalar fields. The
simplest such chiral quark model is the $U(N_f)\times U(N_f)$
linear sigma model (\lsm ).
Let us model each scalar multiplet by a \lsm . If one has only one scalar multiplet one writes  as usual
for a gauged \lsm ,

\be
 {\cal L}(\Sigma ) =
\frac 1 2 \Tr [D_\mu\Sigma D_\mu\Sigma^\dagger]
+\frac 1 2 \mu^2\Tr [\Sigma \Sigma^\dagger] -\lambda \Tr[\Sigma\Sigma^\dagger
\Sigma\Sigma^\dagger]\ + ...
\label{lag0}
\ee

Here   $\Sigma$ contains 9 scalar and 9 pseudoscalar fields. (As usual, $\Sigma$ is a $3\times 3$ complex
matrix, $\Sigma=S+iP= \sum_{a=0}^8(s_a+ip_a)\lambda_a/\sqrt 2$, in which
$\lambda_a$ are the Gell-Mann matrices, normalized as $\Tr[\lambda_a\lambda_b]=
2\delta_{ab}$, and where for the singlet
$\lambda_0 = (2/N_f)^{1/2} {\bf 1}$). If one adds a relatively small
term $\lambda' (\Tr[\Sigma\Sigma^\dagger ] )^2$ the $a_0(980)$ is split from the
$\sigma(600)$. One can add further terms of
higher dimension, and like in Sec. III an anomaly term det$\Sigma +$det$\Sigma^\dagger$ and terms which break the symmetries
explicitely, but these are here not important for our qualitative discussion.

 Thus each  meson in Eq. (2)  belongs to a flavour nonet.
In the quark model this means it can be a $q\bar q$ meson. But,
it need not be $q\bar q$. E.g.,  meson-meson interactions are well known to be
repulsive in exotic channels, but attractive in octet or singlet channels.
Therefore, one expects bound meson-meson states only for octets or singlets.
So we can have another set of states $\hat\Sigma$, which let us assume, are also described  approximately by a similar
Lagrangian as above, $\hat{\cal L} (\hat\Sigma)$, but with another set of parameters ($\hat\mu^2, \hat\lambda$ ).
We have thus doubled the spectrum and  initially we  have  two scalar, and two
pseudoscalar multiplets, altogether 36 states for three flavours.

 Then it is natural to introduce a coupling between the
two sets of multiplets, which can break the relative symmetry.
The full Lagrangian for both $\Sigma$ and $\hat\Sigma$ thus becomes,
\be
 {\cal L}_{tot}(\Sigma ,\hat\Sigma ) ={\cal L}(\Sigma )+\hat{\cal L} (\hat\Sigma)+ \frac {\epsilon^2} 4
 {\rm Tr}[\Sigma\hat\Sigma^\dagger +h.c.]\ .
\label{lag1}
\ee

Without the $\epsilon^2$ term one has for 3 flavours two independent $U(3)\times U(3)$ symmetries ($[U(3)\times U(3)]^2$), but the $\epsilon^2$ term breaks
this down to one
exact over-all $U(3)\times U(3)$ and one broken relative $U(3)\times U(3)$ symmetry.
A similar scheme was discussed recently by Black et al.\cite{black}.

Now let (differently from\cite{black}),
  both $\Sigma$ and $\hat\Sigma$ have  vacuum expectation values\footnote{ On can picture the $U(1)$ or the $O(2)$
 symmetry of this potential (in the case of one flavour) by two coupled "Mexican hat potentials".
 Or, perhaps, better by one Mexican hat (with minimum given by $v$) coupled to another unstable system,
 say a vertical elastic bar which can be  buckled by a strong enough vertical force,
 to the side such that the deflection is $\hat v$. Then when the hat is hanged on the  bar the deflection causes
 the hat to be tilted through the coupling $\epsilon^2$.}
 $v$ and $\hat v$ even if $\epsilon=0$ ($v=\mu^2/(4\lambda)+{\cal O}(\epsilon^2)$,
 $\hat v=\hat \mu^2/(4\hat\lambda)+{\cal O}(\epsilon^2)$.

 Then the $2\times 2$ submatrix between two pseudoscalars with same flavour
becomes to order $\epsilon^2$:
\be
m^2 (0^{-+})=
{\left( \begin{array}{cc}
4\lambda v(\epsilon)^2-\mu^2 & -\epsilon^2  \\
-\epsilon^2& 4\lambda\hat v(\epsilon)^2-\hat\mu^2
 \\
        \end{array}
        \right )
        =+\epsilon^2 \left( \begin{array}{cc}
\hat v/v& -1          \\
-1         & v/\hat v \\
        \end{array}
\right )} \ ,\ee
which is diagonalized by a rotation $\theta$ (tan $(\theta)=  v/\hat v$), such that
the eigenvalues are 0 and $\epsilon^2v\hat v/(v^2+\hat v^2)$:
\be
\left( \begin{array}{cc}
c & -s  \\
s& c \\
        \end{array}\right )m^2 (0^{-+})\left( \begin{array}{cc}

c & s  \\
-s& c \\
        \end{array}\right )=
        \epsilon^2\frac{v^2+\hat v^2}{v\hat v}\left( \begin{array}{cc}
1 & 0  \\
0 & 0  \\
        \end{array}\right )
   \ .      \ee

Here $s=v/\sqrt(v^2+\hat v^2)$ and $c=\hat v/\sqrt(v^2+\hat v^2)$.
The approximation
is of course valid only if neither $v$ nor $\hat v$ vanishes. In fact, one would expect  $v\approx \hat v$
phenomenologically. Thus one has
one massive $|\pi>$ and one massless $|\hat \pi >$ would-be pseudoscalar
multiplet. Denoting the the original pseudoscalars $|p>$ and $|\hat p>$,
we have $|\pi>=c|p>-s|\hat p>$, and $|\hat \pi>=s|p>+c|\hat p>$. The mixing angle is determined entirely
by the two vacuum expectation values,
and is large if $v$ and $\hat v$ are of similar magnitudes, independently of how small $\epsilon^2$ is.
On the other hand the scalar masses and mixings  are only very little affected if $\epsilon^2/(\mu^2-\hat\mu^2)$ is small.
They are still close to $\sqrt 2 \mu $ and $\sqrt 2 \hat \mu$ as in the uncoupled case.

Furthermore, denoting (when $\epsilon=0$) the axial vector current obtained from  ${\cal L}(\Sigma ) $ by $j_{A\mu}$
and the one from ${\cal L}(\hat \Sigma ) $
by $\hat j_{A\mu}$, then their  sum   is exactly conserved ($\partial_\mu [j_{A\mu}+\hat j_{A\mu}]=0$)
because of the masslessness of $|\hat \pi>$, and because of the still exact overall axial symmetry.
But $ j_{A\mu}$ or $\hat j_{A\mu}$ is only "partially conserved",
$\partial_\mu j_{A\mu}=\sqrt{N_f}cs\sqrt(v^2+\hat v^2)m^2_\pi \pi$, because of the $\epsilon^2$ term,
which explicitely breaks the relative symmetry. If one identifies this with PCAC, and $m_\pi$ with the pion mass
then we have

\ba
f_\pi = \sqrt{N_f}v\hat v/\sqrt (v^2+\hat v^2) \ , \\
m_\pi^2=\epsilon^2(v^2+\hat v^2)/v\hat v \ .
\ea

In order that this should have anything to do with reality, one must of course get rid of the massless Goldstones
$\hat \pi$. By gauging the overall axial symmetry i ($D_\mu\Sigma=\partial_\mu\Sigma-i\frac 1 2 g[\lambda_iA_i\Sigma+\Sigma\lambda_iA_i]$).
 I  argue  that the conventional Higgs mechanism absorbs the
massless modes $|\hat \pi>$ from the model, but these degrees of freedom
enter instead as longitudinal axial vector mesons and give these  mesons mass
$m_A^2=2g^2(v^2+\hat v^2)$.
This is similar to the work of Bando et al.\cite{bando} on hidden local symmetries.

The main prediction of this scheme is that one must have doubled the light scalar meson spectrum,
as seems to be experimentally the case. Of course in order to make any detailed comparison with experiment
one must also break the flavour symmetry, and unitarize the model, which is not a simple matter.

The  dichotomic role of the pions in
conventional models, as being at the same time both the Goldstone
bosons and the \qq\ pseudoscalars, is here resolved in a
particularily simple way: One has originally two Goldstone-like pions,
out of which only one remains in the spectrum,
and which is  a particular  linear combination of the two original pseudoscalar fields.

The two scalar multiplets remain as physical states
 and one of these (formed by the $\sigma(600)$ and the $ a_0(980)$ in the case of  two flavours,
 or the $\sigma,\ a_0(980),\ f_0(980)$ and the $ \kappa\  $ in the case of three flavours)
can then be looked upon as the  Higgs
multiplet of strong nonperturbative interactions when a hidden
local  symmetry is spontaneously broken.

\section{Conclusions}

I have in this talk first pointed out that there is strong evidence (Sec. II) for
the existence of the light $\sigma(600)$, and that the linear sigma model (Sec III)
is not an unreasonable approximation for the lightest nonet.
A more detailed understanding requires a unitarized model (Sec IV) whereby one can
undertand how the masses are shifted, the widths and mixings are  distorted, and even why new
scalar meson-meson resonances can be created in nonexotic
channels. These effects are
important  because of the large effective coupling, and because of the nonlinearities due to the S-wave cusps.

The recent observation of $\sigma (600)$ and $f_0(980)$ by the E791 experiment on $D$ and $D_s$
decays to three pions was discussed in Sec. V, and it was pointed out that these rates can be understood
whithin a rather general constituent quark model, with a standard effective Lagrangian for the
weak matrix elements.

At the end, in Sec VI,  I  discussed some preliminary work\cite{fut} concerning
how one could model the the fact that there seems to be
groving experimental evidence for two light nonets of scalars, one in the 500-1000 GeV region
($\sigma(600),$ $ a_0(980),$ $f_0(980),$ $\kappa (?)$),
and another one near 1.3-1.7  GeV ($f_0(1370),$ $ f_0(1500)/f_0(1710),$ $ a_0(1450),$ $ K^*_0(1430)$).
I suggested a linear sigma model as a "toy model" for each nonet, each one with its separate
 vacuum expectation value. Then after coupling these two models
through a mixing term and gauging the overall symmetry, one can argue that one of the nonets
(the lighter one) is a true Higgs nonet for strong interactions.

\section{Acknowledgements}

 Support  from EU-TMR program, contract
CT98-0169 is gratefully acknowledged.

\end{document}